# Violating of the classical Essam-Fisher and Rushbrooke formulas for quantum phase transitions

V. Udodov

Abstract

The classical Essam-Fisher and Rushbrooke relationships (1963) that connect the equilibrium critical exponents of susceptibility, specific heat and order parameter are shown to be valid only if the critical temperature $T_C > 0$ and $T \to T_C$. For quantum phase transitions (PT's) with $T_C = 0K$, these relations are proved to be of different form. This fact has been actually observed experimentally, but the reasons were not quite clear. A general formula containing the classical results as a special case is proposed. This formula is applicable to all equilibrium PT's of any space dimension. The predictions of the theory are consistent with the available experimental data and do not cast any doubts upon the scaling hypothesis.

The critical exponents and scaling hypothesis (Widom (1965), Domb and Hunter (1965); Kadanoff (1966)) underlying the theory of critical phenomena [1-3]. Essam and Fisher equality [4] and Rushbrooke inequality [5] (as the most famous inequality between critical exponents) were discovered 50 years ago and everyone believed that the relations are performed. These relations are connected with the critical exponents of the specific heat (α, α'), susceptibility (γ, γ') and the order parameter $\beta$ (determining of the critical exponents see below) [1-3]. In 1965-66, it was shown that under the scaling hypothesis the Essam and Fisher equality are fulfilled and the Rushbrooke inequality $\alpha' + 2\beta + \gamma' \geq 2$ reduces to equality [1]

$$\alpha' + 2\beta + \gamma' = 2.$$

These results were obtained under the assumption that the critical temperature is finite $T_C > 0$ and $T \to T_C$. However, recently quantum phase transitions (PT's) have inspired a new interest in low-temperature thermodynamics and statistical physics [6-8]. For quantum PT's the temperature of phase transition $Tc$ is zero. In this paper we show that if $T_C = 0K$, the classical Essam-Fisher equation and Rushbrooke inequality changes: the right-hand side equal to two, is replaced by one. Change as well, and some other equations for the critical exponents resulting from the scaling hypothesis. In this paper we propose a general interpolation formula (generalized Essam-Fisher equality), the right side of which ranges from 2 to 1. This general formula is valid for both finite critical temperature and for the case of $T_C = 0K$. In the experimental measurements of the critical exponents for the PT's at low temperatures ($T \to 0$) we can expect that the right side of these relations will be less than 2 [1]. The proposed theory can be tested for any quantum phase transitions [6-9].

Even now there is evidence that these relations are broken. For example, for the magnetic transition in nickel we have [10] (within the framework of the scaling hypothesis $\gamma=\gamma'$)

$$\alpha'+2\beta+\gamma = 1.89 < 2 \ .$$

For the critical point in $CO_2$ [10], for example,

$$\alpha'+2\beta+\gamma' = 1.69 < 2 \ .$$

Explanations of these violations were absent, but everyone believed that the formulas are correct, since derived from thermodynamics. The situation is complicated by the fact that a breach of these relationships speaks about the disturbance of the scaling hypothesis.

It turns out, if $T_C = 0$, the classic relationships have the different form and this may be explains the apparent violations and saves the scaling hypothesis.

First, we recall the usual derived Essam-Fisher equality. The order parameter in a weak field (or lack thereof) is proportional to [1-3]

$$\eta \propto (-t)^\beta \ , \ t < 0 \ , \qquad (1)$$

where $t = T - T_C$, $\beta$ is order parameter critical exponent. On the other hand in a weak field $h$, the order parameter is [3]

$$\eta \approx \chi h \ , \qquad (2)$$

where $\chi$ – susceptibility, the critical exponent is determined by the ratio (in a weak field) [1-3]

$$\chi \propto (-t)^{-\gamma} \ , \qquad (3)$$

$\gamma$ – the critical exponent of the susceptibility. In the field PT becomes fuzzy. If we substitute (1) and (3) in (2), we find blur for temperature interval of PT [3]

$$(-t)^\beta \propto h(-t)^{-\gamma} \ , \qquad (4)$$

from which we obtain

$$h \propto (-t)^{\beta+\gamma} \ . \qquad (5)$$

The same interval of blur can be found from the requirement that a field term of potential $G_h$ is equal to a heat term $G_T$ [3] ($G$ – the Gibbs potential [1-3])

$$G_h \propto \eta h \propto (-t)^\beta h \propto t^2 C_p \propto G_T \ , \qquad (6)$$

where $C_p$ – heat capacity, for which the formula is valid in weak field [1-3]

$$C_P \propto (-t)^{-\alpha'}, t<0; C_P \propto t^{-\alpha}, t>0 , \qquad (7)$$

α, α' – the heat capacity critical exponents, within the framework of the scaling hypothesis α=α' [1]. Substituting (5) and (7) into (6) this gives

$$(-t)^{2\beta+\gamma} \propto (-t)^{2-\alpha} , \qquad (8)$$

or

$$\alpha + 2\beta + \gamma = 2 , \qquad (9)$$

it is Essam and Fisher equality [1, 3, 4].

Longer arguments prove Rushbrooke inequality [1, 5], which is true for the stable states

$$\alpha' + 2\beta + \gamma' \geq 2 , \qquad (10)$$

where the primed symbols correspond to the region below the PT point (at temperature $t=(T-T_C)<0$), the order parameter critical exponent $\beta$ is defined only in this area [1-3].

Note that the heat term has the form (6) only if $T_C > 0$. In fact, let

$$T_C = 0 . \qquad (11)$$

We show that the heat term $G_T$ in the form (6) leads to incorrect results. From (11) it follows that

$$t = T - T_C = T > 0. \qquad (12)$$

The thermal term becomes in the form (within the framework of the scaling hypothesis α=α')

$$G_T = t^2 C_p \propto t^2 t^{-\alpha} \propto t^{2-\alpha} . \qquad (13)$$

We find the entropy

$$S = -\frac{\partial G_T}{\partial T} = -\frac{\partial G_T}{\partial t} \propto (2-\alpha)t^{1-\alpha} \qquad (14)$$

and heat capacity (using (12) and (14))

$$C_P = T\frac{\partial S}{\partial T} \propto t(2-\alpha)(1-\alpha)t^{-\alpha} \propto t^{1-\alpha} , \qquad (15)$$

which contradicts eqn (7).

Let us find the right formula. Let heat term has the form ($T=t>0$, $T_C=0$)

$$G_T = t^b C_p \propto t^b t^{-\alpha} \propto t^{b-\alpha} . \tag{16}$$

We now find $b$. Entropy and heat capacity are

$$S = -\frac{\partial G_T}{\partial T} = -\frac{\partial G_T}{\partial t} \propto (b-\alpha)t^{b-1-\alpha} ,$$

$$C_P = T\frac{\partial S}{\partial T} \propto t(b-\alpha)(b-1-\alpha)t^{b-2-\alpha} \propto t^{b-1-\alpha} \propto t^{-\alpha} , \tag{17}$$

hence we have

$$b = 1$$

and

$$G_T \propto t^{1-\alpha} . \tag{18}$$

Essam and Fisher equality will take the form

$$\alpha + 2\beta + \gamma = 1 \ (T_C = 0) , \tag{19}$$

which differs from the standard formula (9) [1-4].

In the general case may be offered the following interpolation formula (generalized Essam-Fisher equality)

$$\alpha + 2\beta + \gamma = 1 + S_I , \tag{20}$$

where $S$-function is

$$S_I = \left(\frac{T_C}{T}\right)^n, \quad T > T_C , \tag{21}$$

where positive constant $n$ can be found either from the comparison with experiment, or from the microscopic theory. Indeed, if the $T_C = 0$ ($T > 0$), then we arrive at the formula (19). If $T_C > 0$ and $T \to T_C$, we get a classical Essam and Fisher equality (9).

Turns out, if the number $n$ is fraction, then S-function is a nonanalitic function at a point $x = T_C = 0$. For example, if $0 < n < 1$, then the first derivative tends to infinity $(S_I)_x' \to \infty$, $x \to 0$. If $n$ is an integer, then the function $S_I$ can be decomposed into a series of $x$. For $n = 2$, we get decomposition into a series accurate within quadratic term

$$S_I \approx \frac{x^2}{t^2}, \ t > 0, \ \frac{x}{t} = \frac{T_C}{t} << 1 . \tag{22}$$

Now consider the case $T_C = x > 0, t \to 0$. S-function is rewritten as

$$S_I = \left(\frac{1}{1+t/x}\right)^n = \left(\frac{1}{1+\tau}\right)^n, \quad t > 0,$$

where

$$\tau = t/x = \frac{T - T_C}{T_C}, \quad \tau \to 0. \tag{23}$$

Now the S-function is analytic function at a point $\tau = 0$. The decomposition into a series is

$$S_I = 1 - n\tau + \frac{n(n+1)}{2}\tau^2 - ..., \quad \tau \to 0, \tag{24}$$

where $S_I < 1$. Note that (23) and (24) it is also true if $T_C \to 0$, however $\tau \to 0$, it follows from this $T \to 0$.

Rushbrooke inequality in the general case is given by

$$\alpha' + 2\beta + \gamma' \geq 1 + S_I, \tag{25}$$

that is, the right-hand side of (25) can be less than 2. We obtained a principal result: the right side of the classic relationships is not a constant, but rather a function that depends on the temperature and takes the values from 1 to 2.

Finally, the equality

$$\nu d = 2 - \alpha = r, \tag{26}$$

which follows from the scaling hypothesis [1, 3] ($d$ – the space dimension, $\nu$ – critical exponent of the correlation length), also changes its appearance

$$\nu d = 1 + S_I - \alpha = r, \tag{27}$$

$r$ – phase transition order in the sense of Baxter R. [2], we believe that the scaling hypothesis is performed.

It should be noted that the S-function expresses the degree of classical behavior. Thus, $S_I = 0$ for the non-classical behavior and $T_C = 0$ and $S_I = 1$ corresponds to the classical regime.

Thus showed that the classical Essam and Fisher equation and Rushbrooke inequality are only valid if the critical temperature $T_C$ is finite and $T \to T_C$. If the critical temperature is zero, the classic relationships change its form: on the right side instead of the deuce will be unit. Because at change of parameters critical temperature tends to zero for many quantum phase transitions, the right-hand side of these relations should be continuously varied from 2 to 1.

The real test of the results is possible for any quantum phase transitions [7, 8] as $T \to 0$, for example for ferromagnetic $Ce_{2.15}Pd_{1.95}In_{0.9}$ [9] or for heavy-fermions systems [11].

We note that the relations (19), (20), (25) and (27) are equilibrium, which implies that the right side equal to one, is unlikely to be achieved because the relaxation time (the transition to the equilibrium state) at $T \to 0$ tends to infinity.

ACKNOWLEDGMENTS
I thank S. Prosandeev and A. Levanyuk for useful discussions and I am especially grateful for the support to I. Naumov.

REFERENCES

1. Stanley H.E. Introduction to phase transitions and critical phenomena. Physics department Massachusetts institute of technology, Clarendon press. Oxford 1971.
2. Baxter R.J. Exactly Solved Models In Statistical Mechanics / Academic press, London, New York, Sydney, Tokyo, Toronto, 1982.
3. Landau L.D., Lifshitz E.M. Vol. 5. Statistical physics. Part 1. 3ed., Pergamon, 1980.
4. Essam J. W., Fisher M. E., Journ. Chem. Phys., **38**, 802, 1963.
5. Rushbrooke G. S., Journ. Chem. Phys., **39**, 842, 1963.
6. Rechester A.B., Soviet physics JETP, V. 33, No. 2. **60**, 782, 1971. Zh. Eksp. Teor. Fiz. **60**, 782-796, 1971.
7. Sachdev S. Quantum phase transitions. Yale University, New Haven. USA. 1999.
8. Abrikosov A.A. *Fundamentals of the Theory of Metals.* Amsterdam: North-Holland, 1988.
9. Sereni J.G., Giovannini M., Gormez Berisso M., Saccone A. Journal of Physics: Conference Series **391**, 012062, 2012.

10. http://www.chem.utoronto.ca/~hbayat/HanifBayat-Inequalities
11. Stishov S.M., Usp. Fiz. Nauk, **174**, No 8, 853- 860, 2004.